\documentstyle[aps,pre]{revtex}
\begin{document}
\draft
\twocolumn[\hsize\textwidth\columnwidth\hsize\csname @twocolumnfalse\endcsname
\title{An Analytic Description of Light Emission in Sonoluminescence}
\author{Pritiraj Mohanty \\
        {\it Condensed Matter Physics 114-36, California Institute of Technology, 
        Pasadena, CA 91125}  \\
}
\maketitle

\begin{abstract}

Light emission in sonoluminescence is shown to be a lasing process with a wide gain 
bandwidth. Population inversion of the gas molecules inside the bubble is achieved by
hydrodynamical pumping. Analytic expressions are 
derived for the sonoluminescence pulse: shape, 
pulsewidth, and their codependence on the spectrum and intensity in 
physically relevant regimes. A detailed comparison with experiments(R. Hiller 
{\it et al.}, Phys. Rev. Lett. {\bf 80}, 1090 (1998); M.J. Moran
{\it et al.}, Phys. Rev. Lett. {\bf 80}, 4987 (1998); R. Pecha {\it et al.}, 
Phys. Rev. Lett. {\bf 81}, 717 (1998)) suggests excellent agreement. 
 
\vskip 0.1in

\vskip 0.1in
\end{abstract}
]

The phenomenon of sonoluminescence (SL)---emission of a short pulse of light
in an acoustically levitated bubble during its final stage of compression---has 
attracted a lot of interest\cite{rev-son90s}. Unfortunately,  there does not exist 
a complete theory that formally explains all aspects of the emitted 
light\cite{gompf,hiller,moran,pecha}. 
Light emission in SL can be understood 
without invoking bubble dynamics in contrast to the recent 
claims\cite{lohse-nature} that a modified bubble dynamics theory could  
numerically account for the nature of the pulse\cite{gompf,hiller,moran,pecha}. 
We argue that population inversion in the medium of gas molecules, {\it which must 
contain metastable states}, generates the SL pulse; towards that end, we 
explicitly calculate measurable pulse characteristics and compare with the 
experimental observations\cite{hiller,moran,pecha}. However, bubble dynamics 
is important in evaluating the temperature inside the bubble with precision 
from ambient parameters such as pressure, bath temperature and equilibrium radius.

We argue that light emission in SL results\cite{mohanty} mainly due to the following 
two steps: (i) excitation of both gas atoms and molecules containing metastable states
at a rate high enough to sustain population inversion against dephasing and loss, 
(ii) spontaneous and stimulated emission of the excited atomic/molecular levels, 
{\it collective} due to a high density. 
We consider a realistic model of an argon gas bubble and evaluate
various parameters of the emitted light pulse in view of the proposed mechanism of 
sonoluminescence as a broadband  
laser\cite{mohanty}---{\it though an inefficient one due to the presence of strong 
dephasing and loss}. 
In this exercise of SL from an argon bubble\cite{hiller}, we derive explicit analytic 
expressions for (i) the temporal shape of the pulse in various 
relevant regimes, (ii) the pulsewidth consisting of the rise time---denoting 
the growth of the pulse, and the fall time---denoting the decay, both compared with recent 
streak camera experiments in ref.\cite{moran,pecha}, 
(iii) their individual dependence on wavelength, 
compared with experiments in ref.\cite{hiller,moran}, (iv) the intensity of the emitted 
light and its dependence on the pulsewidth, compared with the experiments 
in ref.\cite{hiller}, and (v) the spectrum at different ranges of wavelength, 
compared with experiments in ref.\cite{hiller,hiller-old}. Our results 
are in excellent agreement with  these experiments\cite{gompf,hiller,moran,pecha}.

{\it Chemical Requirement:} It is known that the gas temperature inside an argon gas 
bubble can be as high as 10,000 K to 40,000 K (0.86-3.5 eV)\cite{moss} due to 
hydrodynamic collapse\cite{rev-son90s}. Heating to such a high temperature is sufficient 
to cause electronic excitation of argon gas atoms. This process is necessary in 
creating population inversion in atoms with metastable states essential to standard
lasing process. 
However, for hotter bubbles of argon and other rare gases\cite{helium} 
with higher atomic number (smaller ionization potential) electronically bound 
excited molecular states are formed, 
though the interaction between the 
constituent atoms is mostly repulsive\cite{excimer-book}: 
$\mbox{Ar} + \mbox{Ar}^\ast \rightarrow \mbox{Ar}_2^\ast$, 
which then decay to ground state by emitting a photon: 
$\mbox{Ar}_2^\ast \rightarrow \mbox{Ar} + \mbox{Ar} +h\nu$. 
These excimer states are dissociative 
in the electronic ground state. Spontaneous and stimulated decay of such molecular 
excited states describe the underlying physical mechanism of excimer laser systems 
such as Ar$^\ast_2$ or Xe$^\ast_2$ lasers.
Excimers with strongly repulsive ground states emit broad continuum radiation
whereas the weakly bound excimers typically display a radiation spectrum with
rotational and vibrational structures, which gets homogeneously broadened
at high pressure. The binding energy of the excimers depends
on potential-energy curves which are extremely complex, including families 
of closely nested curves of excited states as well as level crossings linking 
specific levels. After excitation, the system 
evolves from the top to the bottom of these curves by relaxing through a nested 
sequence of electronic states. The chemical 
kinetic ordering originating from the particle interactions of rare gases is 
such that no direct thermal path is available, and the only accessible decay path 
is radiation. We argue that this kind of radiation 
is what gives the light in sonoluminescence.

Let us now consider the chemical kinetics of argon inside the bubble 
specifically for this excercise. During the compression of the bubble at 
high pressures ($P \ge 10^3$ atm) and 
temperatures ($k_BT_e \sim 3.5$ eV),  
argon is excited ($e^{-}+\mbox{Ar} \rightarrow \mbox{Ar}^\ast + e^{-}$) or 
ionized ($e^{-} + \mbox{Ar} \rightarrow \mbox{Ar}^{+} + 2e^{-}$) with a population of
$e^{-E_{ion}/k_BT}$ or $e^{-E_{ex}/k_BT}$ respectively. 
Since the ionization energy of argon is $\sim$ 12 eV, this corresponds to a 
small degree ($\sim 3\%$) of ionization at a temperature in the range of 
0.86-3.5 eV. The atomic excitation to Ar$^\ast$ is much more efficient.
As we will see, a large amount of initial ionization is
not important for the formation of excimer states. 
Three-body recombination reactions rapidly dimerize\cite{keto} the excited
atoms ($\mbox{Ar}^\ast + 2 \mbox{Ar} \rightarrow \mbox{Ar}^\ast_2 + \mbox{Ar}$) or 
ionized states 
($\mbox{Ar}^{+} + 2 \mbox{Ar} \rightarrow \mbox{Ar}^{+}_2 + \mbox{Ar}$)\cite{liu}.
Argon excimers formed mostly in many excited states 
quench to lower states at high pressures\cite{rhodes} and the excitation 
resides mostly in the lowest excited levels of atomic and molecular levels.   
The radiative lifetime $T_1$ of the most relevant excited state $\mbox{Ar}_2^\ast$ 
(for the $1_u$ molecular state which dominates over the $0_u^{+}$ state) is 
known\cite{keto} to be  $3 \times 10^{-6} s$. 

Presence of a small amount ($\ll 10\%$) of a heavier rare gas in a large amount 
of a lighter rare gas\cite{hiller-sci} results in enhanced ionization. These impurities can 
dimerize, recombine and react to form various diatomic gas excimers. If the 
excited species are sufficiently energetic, then they react with the heavier 
rare gas atoms, Xe in Ar for example, 
to form ions in Penning 
($\mbox{Ar}^\ast + \mbox{Xe} \rightarrow \mbox{Ar} + \mbox{Xe}^{+} + e^{-}$) 
or associative ($\mbox{Ar}^\ast + \mbox{Xe} \rightarrow \mbox{ArXe}^{+} + e^{-}$) ionization; 
these reactions are very fast and occur at every collision. 
If Penning or associative ionization is not energetically
favored, then the energy of the lighter species transfers to the heavier species 
by  reactions $\mbox{Ar}^\ast + \mbox{Xe} \rightarrow \mbox{Ar} + \mbox{Xe}^\ast$, 
and $\mbox{Ar}^\ast_2 + \mbox{Xe} \rightarrow 2 \mbox{Ar} + \mbox{Xe}^\ast$\cite{gundel,gleason}.

This roughly describes the processes through which argon gas excimers are formed
inside a sonoluminescing bubble.
In the same spirit, one anticipates other rare gas, rare-gas-halogen excimers and 
quasi-bound metal vapor excimers\cite{excimer-book} to show lasing in the SL 
configuration. The above chemical kinetics analysis applies to all other kinds
of excimers, though the details of the collision mechanisms and reaction rates 
are different. In the following we discuss the second step of the process which 
gives rise to the light emission in SL from an argon bubble: 
decay of the excimer states Ar$^\ast_2$.

{\it  Hamiltonian:}
Let us define a transition operator $R_i^{ab}$ from any state
$|a\rangle$ to $|b\rangle$ of the i-th molecule:
$R_i^{ab}|a\rangle_i = |b\rangle_i$.
The hamiltonian of the molecular system comprising of multiple levels
can then be described by
$H_{at}=\sum_{i,e} \hbar(\omega_0+\omega_e) R_i^{ee} + 
       \sum_{i,g} \hbar\omega_g R_i^{gg}$,
where the index $i$ runs over all the molecules, and the index $e$ and $g$
define the energy of a given excited state $|e\rangle_i$ and ground 
state $|g\rangle_i$ with respect to an average 
energy $\hbar\omega_0$. $\hbar\omega_0$ is chosen, without any loss of generality,
to be the difference in energy between the lowest excited state and the lowest ground
state. 
The hamiltonian of the field is given by
the sum over all modes ${\bf k}$ and polarization $\alpha$ such that
$H_{rad} = \sum_{{\bf k}\alpha} \hbar\omega_{\bf k} a_{\bf k}^{\dagger}a_{\bf k}$.
The interaction term between the molecular states and the field modes is 
$H_{int} = \sum_{{\bf k}egi} \sqrt{\hbar\omega_{\bf k} \over 2\epsilon_0 V}
        [ a_{\bf k} R_i^{ge} d_{ge}(\vec{\bf \epsilon}_\alpha\cdot 
		{\vec{\bf \epsilon}}_{ge}) + h.c.]$.

{\it Polarization and Population Inversion:}
Collective dipole moment of the molecules can be expressed in terms of a 
polarization operator, constructed from the molecular raising and lowering
operators $R^{\pm}_i(\omega)$.
Denoting the projection of the corresponding molecular dipole matrix element $d_{ge}$
along the polarization direction as $d_\omega$, we define the following:
$P^{\pm}({\bf r},\omega,t)= \sum_i d_\omega R_i^{\pm}({\bf r}-{\bf r}_i,\omega,t)$.      
$N ({\bf r},\omega,t) = \sum_i R^{(3)}_i({\bf r}-{\bf r}_i,\omega,t)$
is the number of excited 
molecules at a transition frequency $\omega$.

{\it Maxwell-Bloch Equations:}
The common electric field created by the excited molecules is related to the
polarization by the Maxwell's equation:
\begin{equation}
{\partial^2 {\bf E}^{+}({\bf r},t) \over \partial t^2} -c^2 \nabla \times
\nabla {\bf E}^{+}({\bf r},t) = - {1 \over \epsilon_0} {\partial^2
{\bf P}^{-}({\bf r},t) \over \partial t^2}.
\end{equation}
From the Heisenberg's equation of motion for $ N$ and ${\bf P}$: 
\begin{eqnarray}
\nonumber
{\partial { N} \over \partial t} =
{i\over \hbar} {\bf E}^{-}\cdot[{\bf P}^{+}
 - {\bf P}^{-}] + {i \over \hbar}[{\bf P}^{+}- {\bf P}^{-}]\cdot {\bf E}^{+}\\
{\partial {\bf P}^{+} \over \partial t} = i\omega_0 {\bf P}^{+} +
2i { d^2 \over \hbar} {\vec{\bf \epsilon}}_a \cdot [{\vec{\bf \epsilon}}_a ({\bf E}^{-} {N}
 + {N} {\bf E}^{+})],
\end{eqnarray}
\noindent
which define the Maxwell-Bloch equations(MBE) in operator form\cite{laser-book}. 
Electric field and polarization in an arbitrary direction, the direction 
of the detector $\hat{z}$ for example, can be
expressed in the slow-varying approximation in a complex form as
${\bf E}^{+}({\bf r},t) = \hat{z} {\cal E}^{+}(z,t)e^{i(\omega_0 t -k r)}$ and
${\bf P}^{+}({\bf r},\omega,t) = \hat{z}  {\cal P}^{+}(z,\omega,t) 
e^{-i(\omega_0 t -k r)}$,
where $t$ is replaced by the retarded time $ t-r/c$, 
introduced for convenience, and ${\cal E}$ and ${\cal P}$ are the slow-varying 
envelopes. To describe the experiments it is essential to include the effect of 
dephasing and loss\cite{feld}: $\kappa$ denotes propagation loss of the electric 
field and is related to the refraction index $\eta$ by 
$kc/\omega = \eta -i\kappa$, $1/T_2$ is the dephasing rate of polarization,
$1/T_1$ is the spontaneous emission rate, $\Lambda_p$ is a noise source of polarization 
due to the background field, and finally, $\Lambda$ describes the population inversion 
by the excitation due to hydrodynamic pumping. Eqs. (1) and (2) now take the following form:
\begin{eqnarray}
\nonumber
{\partial{\cal E}^{+} \over \partial z} & = &
    -\kappa {\cal E}^{+} + {i\omega_0 \over 2\epsilon_0 c} 
	\int_{\omega_0-\Delta\omega/2}^{\omega_0+\Delta\omega/2} 
     		{\cal P}^{-}(r,t,\omega) g(\omega) d\omega. \\
\nonumber
{\partial {\cal P}^{+}(r,\omega,t) \over \partial t}  & = &
     \Lambda_p - ({1 \over T_2} - ik v) {\cal P}^{+}
	+ {2 id^2 \over \hbar} {\cal E}^{-} {\cal N}(r,\omega,t), \\
{\partial {\cal N}(r,\omega,t) \over \partial t}  & = &
	\Lambda - {{\cal N}(r,\omega,t) \over T_1}
	- {i \over \hbar}({\cal P}^{+}{\cal E}^{+}- {\cal E}^{-}{\cal P}^{-}).
\end{eqnarray}
\noindent
The first equation describes the electric field created by the polarization
of the molecular excitations of the relevant manifold of states 
minus the loss of the field due to propagation.
$\Lambda_p$ represents the noise polarization due to population inversion and the
blackbody radiation,  
which act as an equivalent input field to stimulate the decay.
The effect of Doppler broadening relevant at high temperatures is 
represented by the term $ikv {\cal P}^{+}$.
In fact, it is not necessary to account for all these phenomenological factors
to describe the essential features of the recent SL experiments.
To that end, we shall obtain the equation for the radiated pulse
for a simplified case in the time domain, by replacing the opeartors
${\cal E}$, ${\cal P}$ and ${\cal N}$ in the MBE by their complex functional
forms. From the first MBE and its complex conjugate, after 
multiplying by 
$c\epsilon_0 A{\cal E}^{\ast}$, an expression for the radiated intensity $I$ is found:
\begin{equation}
{\partial I/\partial z} = -2\kappa I + (\omega_0 A/2) 2 Im[ {\cal E} 
\int {\cal P}(r,\omega,t) d\omega],
\end{equation} 
\noindent
where $A$ is the area of the cross-section perpendicular to the direction $z$.
$\int {\cal P} d\omega$ gives a polarization at a central frequency
$\omega_0$ averaged over an effective bandwidth $(\Delta\omega)_{eff}$.
Differentiating with respect to $t$ and using Eq.~(3) one obtains:
\begin{equation}
{\partial I \over \partial t} = -(1+2\kappa L) {I \over T_2} 
+ {2 \omega_0 A d^2 {\cal N} L\over \hbar} I + \omega_0 AL \Lambda_p {\cal E},
\end{equation}
where $L$ is a characteristic length scale of the active medium. The first term 
on the right hand side describes dephasing or inhomogeneous broadening $(1/T_2)$ 
and propagation loss $(\kappa)$. The second term denotes the source, and it 
can be written as 
${3\lambda_0^2 c \over 4\pi^2}{NL \over T_1} I$. The third term is the polarization 
source for the input field that drives the decay:

\begin{equation}
\Lambda_p {\cal E} = {1 \over \omega_0 A L} {\hbar\omega_0 \over T_R T_2^{eff}}
[ {n_2 \over n_2 -n_1} + {1 \over e^{\hbar\omega_0/k_B T} - 1} ]. \label{input}
\end{equation} 
\noindent
The effective dephasing time\cite{laser-book} for a gain medium is  
$T_2^{eff} = T_2 (2\alpha_0 L/\pi)^{1/2}$, where $\alpha_0 L = T_2/T_R$ is the gain 
over the length $L$, and ${T_R(\omega_0,\Delta \omega_{eff})}^{-1} 
= N(\omega_0,\Delta\omega_{eff})/T_1$  
is the collective decay rate at $\omega_0$.
The second term gives polarization due to blackbody radiation interacting within
$\Delta \omega_{eff}= 1/T_2^{eff}$, and the first term inside the bracket is 
the excitation spectrum of the 
population inversion $g(\omega_0,\Delta\omega_{eff}) = n_2/(n_2-n_1)$. 
In thermal
equilibrium $n_2 = n_1 e^{-\hbar\omega_0/k_B T}$, thus the right hand side of 
Eq. (6) vanishes, implying the absence of population inversion. 

Coupled equations for intensity and population are obtained from Eq. (5) and (6):
\begin{eqnarray}
\nonumber
{\partial I \over \partial t}  =  -(1 + 2\kappa L) {I \over T_2} 
+ {3 \lambda_0^2 c \over 4\pi^2} {NL \over T_1} I \\  
+ {\hbar\omega_0 \over T_R T_2^{eff}} 
	[{g(\omega_0,\Delta\omega_{eff})} + 
		{1 \over e^{\hbar\omega_0/k_B T} - 1} ], \label{rad-intensity}\\
{\partial N \over \partial t}  =  - {N \over T_1^\prime} 
+ \Lambda AL - {2I \over \hbar\omega_0}(1+2\kappa L),
\end{eqnarray}
$T_1^\prime \equiv T_2$ if $T_2 \ll T_1$ as in the case of SL, 
else $T_1^\prime \equiv  T_1$.

{\it Pulse Shape in an Ideal Case:} In an idealistic case\cite{mohanty}
one neglects dephasing, $T_1^{-1},T_2^{-1} = 0 $, 
$\kappa = 0$, and  $\Lambda, \Lambda_p = 0$, retaining the effects
in the initial conditions.
We define a time- and space-dependent tipping angle
$\theta(z,t)$ in terms of the population and polarization in the
medium so that ${\cal N} = n_0 \cos\theta(z,t)$, and ${\cal P}=
i d n_0 \sin\theta(z,t) e^{i\phi}$. The Bloch equations at a given $\omega$ 
reduce to $(2d/\hbar) {\cal E} = e^{i\phi} {\partial \theta/\partial t}$,
and the Maxwell's equation to $\partial^2\theta/\partial z\partial t
= (\Gamma N /L_{\phi}) \sin\theta$, where $\Gamma N= 
T_1 N =  T_R^{-1}$, and $L_{\phi}$ is the
length scale over which the polarization contributes to the 
electric field coherently\cite{preprint}. The MBE can be
written as
\begin{equation}
\theta^{\prime\prime}(u) + \theta^{\prime}(u)/u - \sin\theta =0, \label{sine-gordon}
\end{equation}
\noindent
in a dimensionless variable  
$u = 2 ({z \over L_{\phi}} {t \over T_R})^{1/2}$, demonstrating clearly the 
natural emergence of a collective timescale $T_R$. Solutions to the 
damped pendulum equations of this type, known as sine-gordon equations, include
a wide range of elliptic functions with the constraint 
$\theta(0)=\theta_0$, the input noise field, and $\theta^{\prime}(0)=0$.
If $L \ll L_\phi$, then the entire active medium is coherent and one obtains
pure superradiance.
Solution of the equation in that limit with the first term 
omitted yields an expression for intensity 
$I \propto sech^2(t-T_D/T_R)$, where $T_D = T_R ln(N)$ is the 
pulse rise time\cite{mohanty}. When $z >  L_\phi$, then one expects the occurence of 
several maxima(ringing) in intensity. The field emitted
by molecules in the central coherent volume($z < L_\phi$) are reabsorbed by the 
molecules outside this regime ($ L_\phi < z < R_{min}$) to give stimulated emission. 
Hints of this behavior may have been experimentally observed\cite{moran}.

{\it Pulse Rise Time:}
The pulse rise time is defined as the time at which the first peak is observed from the
time when the sample is completely inverted. In the small angle 
limit ($sin\theta \simeq \theta$), Eq.\ (\ref{sine-gordon}) has the standard Bessel
function solution of the linearized sine-gordon 
equation: $\theta(q) = \theta(0) J_0(iq)$. Using the asymptotic limit, $J_0(iq) \sim
e^{q}/\sqrt{2\pi q}$, and requiring $\theta(q, t = T_D) =1$ for the maximum reached
at time $T_D$, one obtains a transcedental 
expression: $T_D = (T_R/4)[ ln\theta(0)-(1/4)ln(T_R/16\pi^2 T_D) ]^2$. The second term
inside the bracket is negligible compared to the first for the experimentally relevant
parameters, 
and $T_R/16\pi^2 T_D < 10$.
Using Eq. (\ref{input}) one finds $\theta(0) = (2d/\hbar) T_R {\cal E}_{input}
\propto (g(\omega_0) + (e^{\hbar\omega_0/k_B T}-1)^{-1}$. Thus in the range
where the excitation spectrum is thermal, 
($(e^{\hbar\omega/k_B T}-1)^{-1} \gg g(\omega_0))$, 
$T_D = (T_R/4) [ln(k_BT/\hbar\omega)]^2 - \mbox{constant}$, 
for $k_BT \gg \hbar\omega_0$, and 
$T_D = (T_R/4) (\hbar\omega/2k_BT)^2 -\mbox{constant}$, 
for $k_BT \ll \hbar\omega_0$. In the nonthermal regime of the excitation spectrum, 
the rise time is independent of temperature. 
In experiments, since $T$ is usually much larger than $\hbar\omega/k_B$ 
($\sim$ 2500-11000 K in the visible range), dependence of $T_D$ on $\lambda$ and $T$ 
is expected to be 
logarithmically slow, or almost absent, in agreement with the streak camera 
measurements\cite{moran,pecha}.

{\it Pulse Fall Time:} In the limit where dephasing by inhomogeneous broadening 
becomes dominant, $T_2 \ll T_R$, an expression for the radiated intensity 
can be obtained by retaining all the terms in Eq. (\ref{rad-intensity}):
\begin{eqnarray}
\nonumber
I(t)=I(0)e^{-t/\tau_i} 
+ \tau_i {\hbar\omega_0 \over T_R T_2^{eff}} 
   [g(\omega_0,\Delta\omega_{eff}) \\
 + {1 \over {e^{\hbar\omega_0/k_BT} -1}}](1 -e^{-t/\tau_i}),	\label{intensity}
\end{eqnarray}
where the effective fall time is given by 
$\tau_i^{-1} = T_2^{-1}(1+2\kappa L) - 3\lambda_0^2 cL(NT_1^{-1})/ 4\pi^2$. In the absence
of propagation loss ($\kappa \simeq 0)$), and negligible superradiance rate $1/T_R$,
the fall time of the pulse is dominated by dephasing due to inhomogeneous broadening, 
$\tau_i = T_2$.
An expression for $N$ is similarly found to be
\begin{eqnarray}
\nonumber
N(t) = N(0)e^{-t/T_2} + T_1^\prime \Lambda AL(1-e^{-t/T_2})\\ 
- {2I(0) \over \hbar\omega_0} T_1^\prime (1+2\kappa L)(1-e^{-t/T_2})
\end{eqnarray}
\noindent
as $T_1^\prime \simeq T_2 \ll T_1$. Thus in the dephasing dominated regime 
both $I(t)$ and $N(t)$ are seen to decay exponentially with a characteristic time
of $T_2$ rather than with a sech form which is expected in the regime where
dephasing is negligible, $T_2 \gg T_R$. The asymmetry in the pulse shape is
naturally anticipated, since $T_D \ne T_2$. The presence of 
the two timescales and their 
somewhat independent evolution is clearly manifest in the asymmetry observed in 
streak camera experiments(See Figs.~3 and 5 of ref.\cite{pecha}, Fig.~4 of ref.\cite{moran}.)
along with the expected exponential decay of the pulse.

What determines the dephasing time $T_2$? 
Normally at high temperatures $T_2$ is determined by 
Doppler broadening\cite{laser-book}.
The fractional Doppler broadening, $\Delta\omega_d /\omega = 
\sqrt{(8\ln 2)k_BT/Mc^2}$, where $Mc^2$ is the rest mass energy of 
the molecule. Defining the inhomogeneous lifetime\cite{laser-book} as 
$T_2 = 3/\Delta\omega_d$, one obtains, for $\mbox{Ar}_2^\ast$ at T= 20000 K, and 
$\lambda = 400$ nm, $\Delta f_d = 3.75 \mbox{ GHz}$, and a corresponding
$T_2 = 130 \mbox{ ps}$, comparable to the typical measured fall time in the
experiments\cite{hiller,moran,pecha}.
Note that $T_2 \propto 1/\sqrt{T}$ and $T_2 \propto \lambda$,
denoting that Doppler dephasing time is larger for longer wavelengths, which may 
explain why the rise time is observed to be almost independent of $\lambda$ whereas
the fall time increases linearly with $\lambda$ as observed 
in Fig.~3 of ref.\cite{moran}.
But at high pressures, there is a transition from Doppler broadening to 
collisional processes, which depend on the pressure
as well as the collision partner. Correlation experiments
described in ref.\cite{hiller} may be in this regime.

Energy emitted coherently can be obtained by integrating Eq.(\ref{intensity}):
$I(t)=N\hbar\omega/T_R \sin^2\theta e^{-t/\tau_i}$; the maximum
radiated energy is found to be $N\hbar\omega (T_2/T_R)$, indicating that only 
a fraction of the energy is emitted coherently, only for a time $T_2$ shorter 
than $T_R$. The rest of the energy is emitted incoherently\cite{frommhold} 
($T_2/T_R = (\alpha_0 L) \ll 1$).
Thus in the dephasing limited regime one finds that the emitted intensity 
at first linearly increases with increasing pulsewidth, saturating at a value $T_R$
beyond which $T_2$ becomes
irrelevant. The radiated pulsewidth is determined only by $T_R$, as 
observed in the experiments in ref.\cite{hiller}(See for instance Fig.~2 ). 
$T_R$ can be estimated as $\mbox{min}\{T_R(\omega_0)\} \simeq T_1/N(\omega_0)$. A typical 
estimate is
$T_1/N(\omega_0,\Delta\omega_{eff}) = (T_1/N)(f_{max}-f_{min})T_2$. The total 
$N$ of $72 \times 10^7$ is obtained at a density of $100$ times that of STP 
for a volume of $(\pi/6)(\lambda=800 nm)^3$. For the visible range, 
this gives a $T_R$ of 600 ps for $T_1 = 3\times 10^{-6} s$ for 
$\mbox{Ar}_2^\ast$\cite{keto} and  an average $T_2$ of $130 \mbox{ ps}$, 
consistent with experiments\cite{hiller}.

{\it Emitted Spectrum:} 
The energy emitted at a given $\omega_0$ can be obtained from  Eq.~(\ref{intensity}). 
Total energy emitted is found by integrating the energy density 
$(\pi\alpha_0 L/2)^{1/2}[ g(\omega_0)+ (e^{\hbar\omega_0/k_BT}-1)^{-1}]$
over the frequency range with the appropriate
density of states. 
At longer wavelengths (or eqivalently at higher temperatures)
$ (e^{\hbar\omega_0/k_BT}-1)^{-1} \gg g(\omega_0)$, 
thus the emission spectrum is almost thermal
or blackbody-like. This explains why at longer wavelengths (or equivalently in hotter 
bubbles) all the SL spectrums are similar\cite{hiller,hiller-sci,hiller-old}. This is 
not a surprise since the emitted light in this case is actually an amplification 
(with a gain of $\alpha_0 L$) of the input blackbody stimulation. The molecular 
electronic structure is evident at shorter wavelengths when the blackbody term is 
negligible compared to $g(\omega_0)$.    

Dependence of the pulse characteristics on the bath temperature and
ambient pressure can be studied in our model by understanding how they affect
the gas temperature $T$, the intrinsic
parameter in our theory, for which the reader is
referred to the work in ref.\cite{lohse-prl80}.

In conclusion, our model of sonoluminescence as a lasing process with a 
broadband optical gain explains the experimentally observed 
pulse characteristics, the spectrum as well
as the parametric dependences of the emitted light analytically.  
Our calculations and comparisons with experiments will be 
presented in greater detail elsewhere. I thank S.V. Khare and R. Lifshitz 
for many valuable comments.  I acknowledge the formal support of Michael Roukes.



\begin{references}
\bibitem{helium} Lasing from excited molecular states described here does not exclude
the standard lasing from atomic excitation which may dominate in rare gases with
small atomic number, and hence large ionization potential, such
as helium and neon.
\end{references}
\end{document}